\documentclass[twocolumn,superscriptaddress,floatfix,nofootinbib,prx]{revtex4}

\usepackage{graphicx}
\usepackage{mathrsfs}
\usepackage{amsmath}
\usepackage{amsfonts}
\usepackage{subfigure}
\usepackage{bm}
\usepackage{verbatim}
\usepackage{color}
\usepackage{xcolor}
\usepackage[colorlinks=true, letterpaper=true, pdfstartview=FitV, linkcolor=blue, citecolor=blue, urlcolor=blue]{hyperref}
\usepackage{siunitx}
\usepackage{braket}
\usepackage{physics}
\usepackage{float}

\hypersetup{%
pdftitle={Excitonic Laughlin States in Ideal Topological Insulator Flat Bands and Possible Presence in Moiré Superlattice Materials},%
pdfauthor={Nikolaos Stefanidis and Inti Sodemann},%
pdfpagemode={UseNone},%
pdfstartview={FitH},%
breaklinks=true}

\newcommand{\be}{\begin{equation}}
\newcommand{\ee}{\end{equation}}
\newcommand{\ba}{\begin{equation}\begin{split}}
\newcommand{\ea}{\end{split}\end{equation}}

\begin{document}
\title{Excitonic Laughlin States in Ideal Topological Insulator Flat Bands and Possible Presence in Moir\'e Superlattice Materials}
\author{Nikolaos Stefanidis}
\author{Inti Sodemann}
\affiliation{Max-Planck Institute for the Physics of Complex Systems, D-01187 Dresden, Germany}

\begin{abstract}
We investigate few- and many-body states in half-filled maximally symmetric topological insulator flat bands realized by two degenerate Landau levels which experience opposite magnetic fields. This serves as a toy model of flat bands in moir\'e materials in which valleys have Chern numbers $C= \pm 1$. We argue that although the spontaneously polarized Ising Chern magnet is a natural ground state for repulsive Coulomb interactions, it can be in reasonable energetic competition with correlated states which can be viewed as Laughlin states of excitons when short distance corrections to the interaction are included. This is because charge neutral excitons in these bands behave effectively as charged particles in ordinary Landau levels. In particular, the Ising Chern magnet is no longer the ground state once the strength of a short range intra-valley repulsion is about $30\%$ larger than the inter-valley repulsion. Remarkably, these  excitonic Laughlin states feature valley number fractionalization but no charge fractionalization and a quantized charge Hall conductivity identical to the Ising magnet, $\sigma_{xy}= \pm e^2/h$, and thus cannot be distinguished from it by ordinary charge transport measurements. The most compact excitonic Laughlin state that can be constructed in these bands is an analogue of $\nu=1/4$ bosonic Laughlin state and has no valley polarization even though it spontaneously breaks time reversal symmetry with a charge Hall conductivity $\sigma_{xy}= \pm e^2/h$.
\end{abstract}

\maketitle

\noindent

\section{Introduction}

The most experimentally generous platform to realize fractionalized phases of matter to this date are partially filled low Landau levels of clean two-dimensional electron systems subjected to strong quantizing magnetic fields. Landau levels are essentially flat bands with a non-zero Chern number \cite{Thouless82}. The facility towards fractionalization in Landau levels is intimately tied to such band topology which obstructs the construction of  a complete set of localized orbitals  \cite{Thouless84,Thonhauser06}. The latter forbids the existence of a meaningful Hubbard-type limit in which interactions are diagonal in the orbital position basis and devoid from quantum fluctuations, forcing any physically realistic model of interactions to always lead to strong quantum fluctuations in the location of the particles even in the flat band limit, and ultimately leading to the melting of charge-density-wave states in favor of correlated Laughlin-type states over a wide range of conditions. 

Recently a new platform for the appearance of extremely flat bands with non-trivial topology has emerged in the form moir\'e super-lattices of two-dimensional materials. Following the prediction of the appearance flat bands at small twist-angle in graphene moiré super-lattices \cite{Bistritzer11,Morell10}, experiments have observed a variety of superconducting and correlated insulating states in these \cite{Cao181,Sharpe19,Cao182,Yankowitz19,Choi19,Lu19,Xie2019,Wong19} and other moiré superlattice materials \cite{Chen191,Chen19,Shen20,Cao19,Burg19,Tang19,Vries20}.
\begin{figure}
\begin{center}
\hspace{-0.2in}
\includegraphics[width=3.4in]{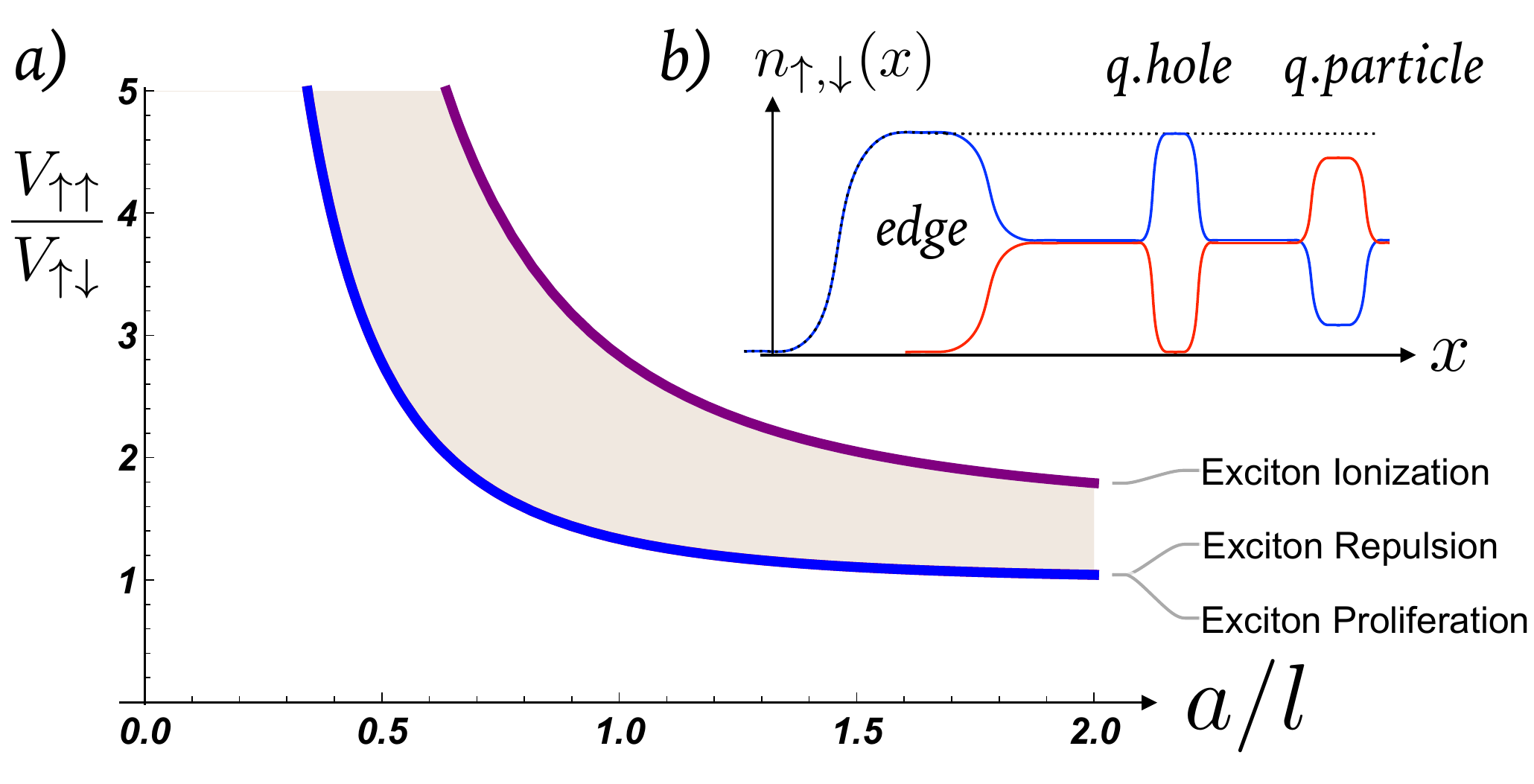}
\caption{(a) Shaded region of optimal conditions for excitonic Laughlin states for interactions with range $a$ and intra- and inter-valley strengths $V_{\uparrow \uparrow}$ and $V_{\uparrow \downarrow}$.  Above blue line: excitons proliferate and repel. Below purple line: exciton binding energy is larger than inter-exciton interaction. (b) Schematic of the most compact exciton Laughlin state with zero valley polarization but Hall conductivity identical to Chern magnet $\sigma_{xy}=\pm e^2/h$. Charge neutral valley fractionalized quasiparticles are depicted. Red and blue line are electron densities of each valley and dotted line is the total density.}
\label{Fig1}
\end{center}
\end{figure}
The mechanisms behind these phenomena have been strongly debated \cite{SongBernevig2019,Xu2018,Volovik2018,Yuan2018,Po2018,Baskaran2018,Dodaro2018,Padhi2018,
XuLee2018,LiuZhang2018,Isobe2018,Thomson2018,Kang2019,Slagle2020,
ZhangSenthil2020,KhalafVishwanath2020}. Moreover super-lattices aligned with a hexagonal boron nitride (hBN) substrate can harbour valleys with opposite Berry curvatures \cite{Song15} and can lead to moir\'e mini-bands in which the two valleys have flats bands with opposite Chern numbers \cite{Zhang19,Liu19,Chittari19,Bultinck19,Zhang192} and wavefunctions of the continuum model have been argued to be reminiscent of those in Landau levels \cite{Liu19,Tarnopolsky19,BultinckKhalaf19}. In fact, experiments have observed hysteretic \cite{Sharpe191} and quantized anomalous Hall effect in twisted bilayer graphene \cite{Serlin20} and trilayer graphene \cite{Chen192} moir\'e super-lattices on hBN substrates. These states have been theoretically rationalized as spontaneously valley polarized magnetic Chern insulators \cite{Zhang19,Bultinck19,Repellin19,Wu20,Chatterjee19,Alavirad19,YiZhang20,Xie20,Liu192}. 

In the present study, we take an idealized limit of these systems in which the valleys are literally viewed as two Landau levels experiencing opposite magnetic fields, and the physical spin is taken to be fully polarized. This can also be viewed as a maximally symmetric model of a topological insulator flat band \cite{Bernevig96}. We advance several results on the few- and the many-body problems in these systems. We show that pairs of electrons with opposite valleys behave like excitons in the usual quantum Hall context, in the sense that their momentum is locked to be proportional and orthogonal to the relative distance between the particles. Conversely charge neutral inter-valley excitons will be shown to behave like a pair of charged particles in a magnetic field. We will also demonstrate that the Chern magnet is a stable exact ground state for a large class of repulsive Hamiltonians and study its stability against single particle, exciton and exciton pair proliferation instabilities. In spite of its stability, we will argue that there are competing correlated states in the form of Laughlin states of excitons, which could be stabilized with moderate modifications of the relative strength of short-distance intra- and inter-valley repulsive interactions. Specifically, these states are expected to be energetically competitive once the following three conditions are satisfied: (a) The Ising Chern magnet is unstable against exciton proliferation, (b) the excitons remain strongly bound, and (c) the inter-exciton interactions are repulsive. In the shaded region in Fig.\ref{Fig1}(a) all these three conditions are satisfied for a toy model with Gaussian repulsive interactions described in Eq.~\eqref{G}. Remarkably, these states have only valley fractionalization but no charge fractionalization and their Hall conductivity is expected to be identical to that of the ordinary Ising Chern magnet ($\sigma_{xy}= \pm e^2/h$), as depicted in Fig.\ref{Fig1}(b) making it hard to distinguish them via conventional charge transport experiments. In particular, we will show that the most compact excitonic Laughlin state, which is presumably also the most stable one, has zero valley polarization in spite of spontaneously breaking the time reversal symmetry with a charge Hall conductivity of $\sigma_{xy}= \pm e^2/h$. Notice that the excitonic Laughlin states are sharply distinct from the more conventional analogues of Laughlin states studied in fractionally filled chern bands of moiré materials \cite{Ledwith19,Repellin19Senthil,Ahmed20}.

\noindent
\section{Maximally Symmetric 2D Topological Insulators}Our primary interest is to study the ground states of electrons partially filling the flat bands of a time-reversal invariant topological insulator that interact via repulsive forces. To do so it is convenient to consider the largest symmetry that is compatible with such topological insulator band topology. This will allow us to simplify substantially the understanding and the construction of correlated states, and it is a natural starting point before adding realistic perturbations that break these symmetries, such as the band dispersion. In a sense, to not follow this path would be like trying to understand correlated states in Chern bands before understanding them in the ideal isolated Landau levels. 
\par A maximally symmetric realization of a topological insulator band is comprised of two Landau levels with opposite Chern numbers, $C=\pm 1$~\cite{Bernevig96}. We can view them as the $n=0$ Landau level of a Hamiltonian for particles in two valleys that experience opposite magnetic fields:
\begin{equation}\label{H}
H=\frac{(\mathbf{p}-\sigma_{z}\mathbf{A})^2}{2m},
\end{equation}
here $\sigma_{z}= \pm 1$ would be the valley index of two bands with opposite Chern number in the corresponding moir\'e super-lattice material, $\mathbf{A}$ is the vector potential for a uniform perpendicular magnetic field: $\nabla \times \mathbf{A}=B \mathbf{\hat{z}}$, and the magnetic field would be interpreted simply as the scale that controls the area of the moir\'e super-lattice unit cell, $a_{UC}=2\pi l^2= 2\pi/B$. In addition we will take the particles to experience valley dependent interactions that separately preserve the number of particles in each valley, of the form:
\begin{equation}\label{V}
V=\sum_{i<j}V_{\uparrow \uparrow}(\mathbf{r}_i-\mathbf{r}_j)\delta_{\sigma_{zi},\sigma_{zj}}+V_{\uparrow \downarrow}(\mathbf{r}_i-\mathbf{r}_j)\delta_{\sigma_{zi},-\sigma_{zj}},
\end{equation}
where $V_{\uparrow \uparrow}$ and $V_{\uparrow \downarrow}$ denote intra- and inter-valley interactions respectively. The physical Hamiltonian is understood to be the interaction from Eq.~\eqref{V}  projected onto one of the doubly degenerate Landau levels defined by Eq.~\eqref{H} which we will take to be the $n=0$ for concreteness.
\par The projected Hamiltonian enjoys a large symmetry group. In particular, it has three spatially local symmetries that will be crucial in our subsequent analysis. These are the $U(1)\times U(1)$ valley resolved particle number conservation:
\begin{equation}\label{U}
Uc^{\dagger}_{m\sigma}U^{\dagger}=e^{i\phi_{\sigma}N_{\sigma}}c^{\dagger}_{m\sigma},U=e^{i\sum_{\sigma}\phi_{\sigma}N_{\sigma}},
\end{equation}
where $c^{\dagger}_{m\sigma}$ is the electron creation operator for valley $\sigma$ and intra-Landau level index $m$, and $\phi_{\sigma}$ is a valley dependent U(1) phase. The anti-unitary time reversal symmetry ($T^2=-1$):
\begin{equation}\label{T}
Tc^{\dagger}_{m\sigma}T^{-1}= i\sigma^{y}_{\sigma,\sigma'}c^{\dagger}_{m\sigma'},T=i\sigma^{y}K.
\end{equation}
And a unitary charge conjugation symmetry, that maps particles into holes of the opposite valley, and can be chosen as:
\begin{equation}\label{C}
Cc^{\dagger}_{m\sigma}C^{\dagger}= i\sigma^{y}_{\sigma,\sigma'}c_{m\sigma'}.
\end{equation}
We note that the valley conservation and time reversal can be enforced in systems with and without boundaries, but the particle-hole conjugation can only be strictly enforced in geometries without boundaries, such as the sphere or the torus, and boundaries will induce particle-hole symmetry breaking terms. Additionally, the projected Hamiltonian has a rich space symmetry group, which is larger than an ordinary lattice Hamiltonian. For example, in infinite space, it is endowed with a continuous magnetic translational algebra, analogous to the magnetic translation algebra of Landau levels, and whose single particle generators can be written as:
\begin{equation}\label{Q}
\mathbf{Q}=\boldsymbol\pi+\sigma_{z}B\mathbf{\hat{z}}\times \mathbf{r}, 
\end{equation}
where $\boldsymbol\pi=\mathbf{p}-\sigma_z \mathbf{A}$ is the mechanical momentum operator, which satisfies the commutation relations:
\begin{equation}
[Q_i,Q_j]=-i \sigma_z B \epsilon_{ij}.
\end{equation}
The closely related projected position operator defines an analogue of the guiding center operators of Landau levels, and are given by:
\begin{equation}\label{R}
\mathbf{R}=-\frac{\sigma_z}{B}\mathbf{\hat{z}}\times\mathbf{Q}=
\mathbf{r}-\frac{\sigma_z}{B}\mathbf{\hat{z}}\times\boldsymbol\pi.
\end{equation}
The translation operator satisfies therefore a non-commutative algebra that encodes a valley dependent Aharonov-Bohm effect of the form:
\begin{equation}
t_{\mathbf{a}}t_{\mathbf{b}}=t_{\mathbf{b}}t_{\mathbf{a}}e^{i \sigma_z B \mathbf{\hat{z}} \cdot(\mathbf{a}\times \mathbf{b})}, 
\end{equation}
where $t_{\mathbf{a}}=e^{-i \mathbf{a}\cdot \mathbf{Q}}$ and $\mathbf{a}$ is 2D vector. 

In fact, this many body translational magnetic symmetry can be used to rigorously prove that valley unpolarized states and time reversal invariant states do not have exact topological degeneracies, in contrast to usual Landau levels~\cite{Haldane85}, as we demonstrate in the Appendix \ref{Appendix}.

\section{Two-particle problem} Before tackling the fully fledged complexities of the many-body problem, we will begin by analyzing the two-particle problem. This problem is of great significance in conventional Landau levels, because it can be solved essentially by exhausting the symmetries of the problem, and in particular, the center of mass position and the relative angular momentum serve to label unique two-body states and their energy defines the useful notion of Haldane pseudopotentials~\cite{Prange90}.
As we will see, however, the two-body problem for flat topological insulator bands can also be fully solved by employing symmetries, but its structure is very different from that of usual Landau levels and will resemble rather the problem of two particle of opposite charge (neutral exciton) in conventional Landau levels~\cite{Gorkov67,Halperin84}, as it is intuitively clear from the form of the Hamiltonian in Eq.~\eqref{H}.

\par Let us define guiding center relative distance ($\mathbf{d}$) and center of mass position ($\mathbf{R}$) for particles $1$ and $2$ as follows:
\begin{equation}
\mathbf{R}=\frac{\mathbf{R_1}+\mathbf{R_2}}{2},\ \mathbf{d}=\mathbf{R_1}-\mathbf{R_2},
\end{equation} 
where $\mathbf{R_1},\mathbf{R_2}$ are defined in Eq.~\eqref{R}. These operators satisfy the following commutation relations:
\begin{equation}
\begin{split}\label{d}
[d_i,d_j]&=4[R_i,R_j]=-\frac{i \ (\sigma_{z1}+\sigma_{z2}) \ \epsilon_{ij}}{B}, \\
[d_i,R_j]&=-\frac{i \ (\sigma_{z1}-\sigma_{z2}) \ \epsilon_{ij}}{2B}.
\end{split}
\end{equation}
The closely related magnetic center of mass ($\mathbf{Q}$) and relative momentum ($\mathbf{q}$), can be defined as:
\begin{equation}
\mathbf{Q}=\mathbf{Q_1}+\mathbf{Q_2},\ \mathbf{q}=\frac{\mathbf{q_1}-\mathbf{q}_2}{2},
\end{equation}
where $\mathbf{Q_{1,2}}$ are defined in Eq.~\eqref{Q}, and satisfy the following commutation relations:
\begin{equation}\label{17}
\begin{split}
&[Q_i,Q_j]=4[q_i,q_j]=-i(\sigma_{z1}+\sigma_{z2})B\epsilon_{ij}, \\
&[Q_i,q_j]=-i\frac{(\sigma_{z_1}-\sigma_{z_2})B \epsilon_{ij}}{2}, \\
&[R_i,Q_j]=[d_i,q_j]=i \delta_{ij}, \\
&[d_i,Q_j]=[R_i,Q_j]=0. \\
\end{split}
\end{equation}
\par Let us now exploit the symmetries of the Hamiltonian in Eq.~\eqref{H} for the case of two particles. First the valley pseudo-spin of each particle, $\sigma_{z_1},\sigma_{z_2}$, is a conserved number. In the case of valley polarized states in which the two-particles have the same pseudo-spin, $\sigma_{z_1}=\sigma_{z_2}$, the algebra
reduces to that of conventional Landau levels and the problem can be solved following the standard approach via Haldane pseudo-potentials~\cite{Prange90}. We will therefore focus on the case of valley un-polarized states: $\sigma_{z_1}=-\sigma_{z_2}$. We will take the particles to be distinguishable and the desired wavefunctions of bosons and fermions can be obtained from our results by performing the corresponding symmetrization or antisymmetrization. Now, in this case the Hamiltonian between the two particles is only a function of the relative coordinate variable:
\begin{equation}
V^{P}_{\uparrow \downarrow}= P_{0}V_{\uparrow \downarrow}(\mathbf{r_1}-\mathbf{r_2})P_{0},
\end{equation}
where $P_{0}$ denotes projection onto the valley degenerate Landau levels. From Eq.~\eqref{d} we see that for $\sigma_{z_1}=-\sigma_{z_2}$, the relative coordinates along both directions commute and can be simultaneously diagonalized:
\begin{equation}
[d_x,d_y]=0,\ \mbox{for} \ \sigma_{z_1}=-\sigma_{z_2}.
\end{equation}
Therefore the eigenfunctions can be parametrized by a vector of two real numbers $\mathbf{d}=(d_x,d_y) \in \mathbb{R}^2$, and can be written formally as:
\begin{equation}
\ket{\psi_{\uparrow \downarrow}(\mathbf{d})}=\ket{d_x,d_y}_{12}\otimes \ket{\sigma_{z}}_1\ket{-\sigma_{z}}_2.
\end{equation}
These states will have an eigen-energy which is simply a convolution of the unprojected interaction potential and the form factor squared of the Landau level of interest:
\begin{equation}
E_{\uparrow \downarrow}(\mathbf{d})=\int \frac{d^2 q}{(2 \pi)^2}V_{\uparrow \downarrow}(\mathbf{q})|F(\mathbf{q})|^2 e^{i \mathbf{q}\cdot \mathbf{d}}.
\end{equation}
For example, if the unprojected interaction is a delta function $V_{\uparrow \downarrow}=g_{\uparrow \downarrow} \delta(\mathbf{r})$, and we project into the $n=0$ Landau level of Galilean fermions, $|F_{0}(\mathbf{q})|^2=e^{-\frac{l^2|\mathbf{q}|^2}{2}}$, then the energy will be $E_{\uparrow \downarrow}(\mathbf{d})=g_{\uparrow \downarrow} e^{-\frac{|\mathbf{d}|^2}{2l^2}}/2 \pi$.
Notice that although the two particles have a well defined separation vector $\mathbf{d}$, they do not have a well average or center of mass position $\mathbf{R}$, since Eq.~\eqref{d} implies that these two sets of variables cannot be simultaneously specified when $\sigma_{z_1}=-\sigma_{z_2}$. The particles, however, have a well defined total or center of mass magnetic momentum $\mathbf{Q}$, which can be simultaneously specified together with $\mathbf{d}$, according to Eq.~\eqref{17}. However, this is not an independent variable but it is locked to be proportional and orthogonal to $\mathbf{d}$, as follows:
\begin{equation}
\mathbf{Q}=-\sigma_{z_1}B\mathbf{\hat{z}} \times \mathbf{d},\ \mbox{for} \ \sigma_{z_1}=-\sigma_{z_2}.
\end{equation}
Therefore, we see that the behavior of a pair of charged particles in the flat topological insulator band resembles that of a neutral particle-hole pair in ordinary Landau levels.

\section{Ising Chern Magnets: ideal Hamiltonians and exact ground states}
In this section we will exploit the individual valley particle number conservation to demonstrate that the spontaneous Ising Chern magnetic insulator that appears at total filling $\nu=1$ is the exact many-body ground state for a wide class of repulsive Hamiltonians.

The valley conservation described in Eq.~\eqref{U} allows to separate the Hilbert into subspaces labeled by the number of particles in each valley $(N_{\uparrow},N_{\downarrow})$.  Let us consider the system to be placed in a finite geometry, such as torus, which restricts $0\leq N_{\uparrow, \downarrow}\leq N_{\phi}$. Then, there are four subspaces that have a single state, and therefore are automatically guaranteed to be exact eigenstates of any Hamiltonian with valley number conservation, namely $(0,0), (0,N_{\phi}), (N_{\phi},0), (N_{\phi},N_{\phi})$. The states $\Psi(0,0),\Psi(N_{\phi},N_{\phi})$ are the completely empty and the completely filled topological insulator band, whereas $\Psi(0,N_{\phi}),\Psi(N_{\phi},0)$ represent fully polarized magnetic Chern insulators.

\par Although enforcing valley conservation is a useful theoretical device, this is never an exact symmetry in experiments. Additionally, even though it is easy to fix the total particle number, $N_{\uparrow}+N_{\downarrow}$, experimentally it is much harder to imagine a experimental knob that would control the valley polarization $N_{\uparrow}-N_{\downarrow}$. Therefore, it is natural to consider the problem of the absolute ground state of the problem at fixed total particle number, $N_{\uparrow}+N_{\downarrow}$, for the various allowed valley polarizations. In particular, in this section we will concentrate in determining when the fully polarized Ising magnets $\Psi(0,N_{\phi}),\Psi(N_{\phi},0)$ are the absolute ground states at total particle $N_{\uparrow}+N_{\downarrow}=N_{\phi}$. These two states are exchanged either by the time reversal symmetry or the charge conjugation symmetry $T$ and $C$ defined in Eqs.~\eqref{T},~\eqref{C} and are therefore degenerate whenever one of these symmetries is enforced, in which case they would break either of these symmetries spontaneously. We will now introduce a set of ideal Hamiltonians for which these states are the exact ground states at $N_{\uparrow}+N_{\downarrow}=N_{\phi}$. Consider the following interacting Hamiltonian: 
\begin{equation} \label{23}
V_{0}= \sum_{i<j} V_{\uparrow \uparrow} l^2 \delta^{(2)}(\mathbf{r_i}-\mathbf{r_j}) \delta_{\sigma_{zi},\sigma_{zj}} +  V_{\uparrow \downarrow}(\mathbf{r_i}-\mathbf{r_j})\delta_{\sigma_{zi},-\sigma_{zj}}.
\end{equation} 
The above Hamiltonian is understood to be projected onto one of the doubly degenerate Landau levels of Eqs.~\eqref{H}. Here we demand the  interactions to be strictly repulsive, namely:
\begin{equation} \label{24}
V_{\uparrow \uparrow}>0,\ V_{\uparrow \downarrow}(\mathbf{r})>0 \ \forall \ \mathbf{r}.
\end{equation}
The above Hamiltonian is a positive semidefinite operator, namely its exact eigenstates have energies $E \geq 0$. Now, we can see that the Ising magnets are exact zero energy ground states of this Hamiltonian:
\begin{equation}
V_{0}\Psi(0,N_{\phi})=V_{0}\Psi(N_{\phi},0)=0.
\end{equation}
The above follows from the fact that the Pauli exclusion principle guarantees that particles of the same valley are never at the same spatial location and therefore the delta function part of the intra-valley interaction is always identically zero for any fermionic wavefunction~\cite{footnote}.
Now let us consider a partially polarized eigenstate $(N_{\uparrow},N_{\downarrow})$. This state will have an inter-valley correlation function of the form $g_{\uparrow \downarrow}(\mathbf{r_{\uparrow}},\mathbf{r_{\downarrow}})$, which we take to measure the probability density of finding a particle of valley pseudo-spin $\sigma_{z}$ at position $\mathbf{r_{\downarrow}}$ and a particle of valley pseudo-spin $-\sigma_{z}$ at position $\mathbf{r_{\downarrow}}$, given by:
\begin{equation} \small
\begin{split}
&g_{\uparrow \downarrow}(\mathbf{r_{\uparrow}},\mathbf{r_{\downarrow}})= \\
& \Psi^{\dagger}(N_{\uparrow}, N_{\downarrow})\Big(\sum_{i,j} \delta^{2}(\mathbf{\hat{r}}_i-\mathbf{r_{\uparrow}})\delta^{2}(\mathbf{\hat{r}}_j-\mathbf{r_{\downarrow}})\delta_{\sigma_{zi},-\sigma_{zj}})\Big)\Psi(N_{\uparrow}, N_{\downarrow}).
\end{split}
\end{equation}
Now the above quantity is always non-negative $g_{\uparrow \downarrow}$, but moreover, if the state $\Psi(N_{\uparrow}, N_{\downarrow})$ is partially valley polarized, this function will be strictly positive at least for some region of finite measure. This is because if there is a particle of say valley $\sigma_z=1$ at $\mathbf{r_{\uparrow}}$ there is some finite probability to find another particle at some other location $\mathbf{r_{\downarrow}}$ with $\sigma_{z}=-1$. Now, because the energy of the eigenstate in question can be written as:
\begin{equation}
E_{\Psi}=\frac{1}{2}\int d\mathbf{r_{\uparrow}} d\mathbf{r_{\downarrow}} g_{\uparrow \downarrow}(\mathbf{r_{\uparrow}},\mathbf{r_{\downarrow}}) V_{\uparrow \downarrow}(\mathbf{r_{\uparrow}},\mathbf{r_{\downarrow}}),
\end{equation}
it is clear that the condition stated in Eq.~\eqref{24} guarantees the energy of any partially polarized state $\Psi(N_{\uparrow}, N_{\downarrow})$ to be strictly positive $E_{\Psi}>0$. In fact, due to the inability to make fully localized single-particle wavefunctions on a Landau level, it is highly likely that for partially polarized wavefunction $\Psi(N_{\uparrow}, N_{\downarrow})$ the function $g_{\uparrow \downarrow}$ can only be made zero in a subset of measure zero, for example one could have very high degree zero in a correlated wavefunction when particles approach each other $g_{\uparrow \downarrow}(\mathbf{r_{\uparrow}},\mathbf{r_{\uparrow}}\rightarrow \mathbf{r_{\downarrow}}) \rightarrow 0$. Therefore, to guarantee the strict positivity of the energies of partially polarized states, $E_{\Psi}>0$, it is sufficient to demand that $V_{\uparrow \downarrow}(\mathbf{r})>0$ only for $\mathbf{r}$ belonging to some region of the $2D$ plane with non-zero measure, for example simply by being non-zero inside a finite radius defining a hard-core. With this we conclude the rigorous arguments showing that the spontaneously polarized Ising Chern magnets are exact unique zero energy ground states of the class of Hamiltonians introduced in Eq.~\eqref{23}.
\par Now, when the intra-valley interactions are not delta functions, the Ising magnets $\Psi(0,N_{\phi}),\Psi(N_{\phi},0)$ can still be the absolute ground states, even though it is harder to make rigorous statements in this case. Of particular interest for moir\'e superlattice materials is the case when intra-valley interactions and inter-valley interactions are identical. The reason is that because the orbitals at different valleys are related by the microscopic physical time reversal symmetry, which is local in real space. Therefore the probability amplitude of orbitals in both valleys will have the same space structure, and the leading density-density interactions between particles in the same and opposite valleys are therefore expected to be the same. As we will see in the coming sections, there are good reasons to expect that the Ising Chern magnets remain as robust unique ground states for a larger class of such repulsive Hamiltonians, even though it is harder to make rigorous proofs.

\section{Excitons and stability of Ising Chern magnets} \label{Excitons and stability of Ising Chern magnets}
Having established that the Ising Chern magnet is an exact and unique ground state for a wide class of repulsive Hamiltonians, we will now study the excitations on these states. To do so, it is convenient to
perform a partial particle-hole transformation of the fully occupied flavor, which we take to be $\uparrow$, as follows:
\begin{equation} \label{P}
P c^{\dagger}_{m \uparrow} P^{\dagger}=c_{m \uparrow}, \ P c^{\dagger}_{m \downarrow} P^{\dagger}=c^{\dagger}_{m \downarrow}.
\end{equation}
This transformation reverses the sign of the Chern number of valley $\uparrow$, allowing to view the system as a type quantum Hall bilayer in which both flavors experience the same magnetic field at the expense of reversing the sign of the inter-valley interaction and making it effectively attractive~\cite{Zhang18}. We will refer to the representation of the problem
after this transformation as the “quantum Hall” picture, whereas the original representation will be referred to as the “topological insulator” or “physical” picture.
\begin{figure}[h!]
\begin{center}
\includegraphics[width=3.4in]{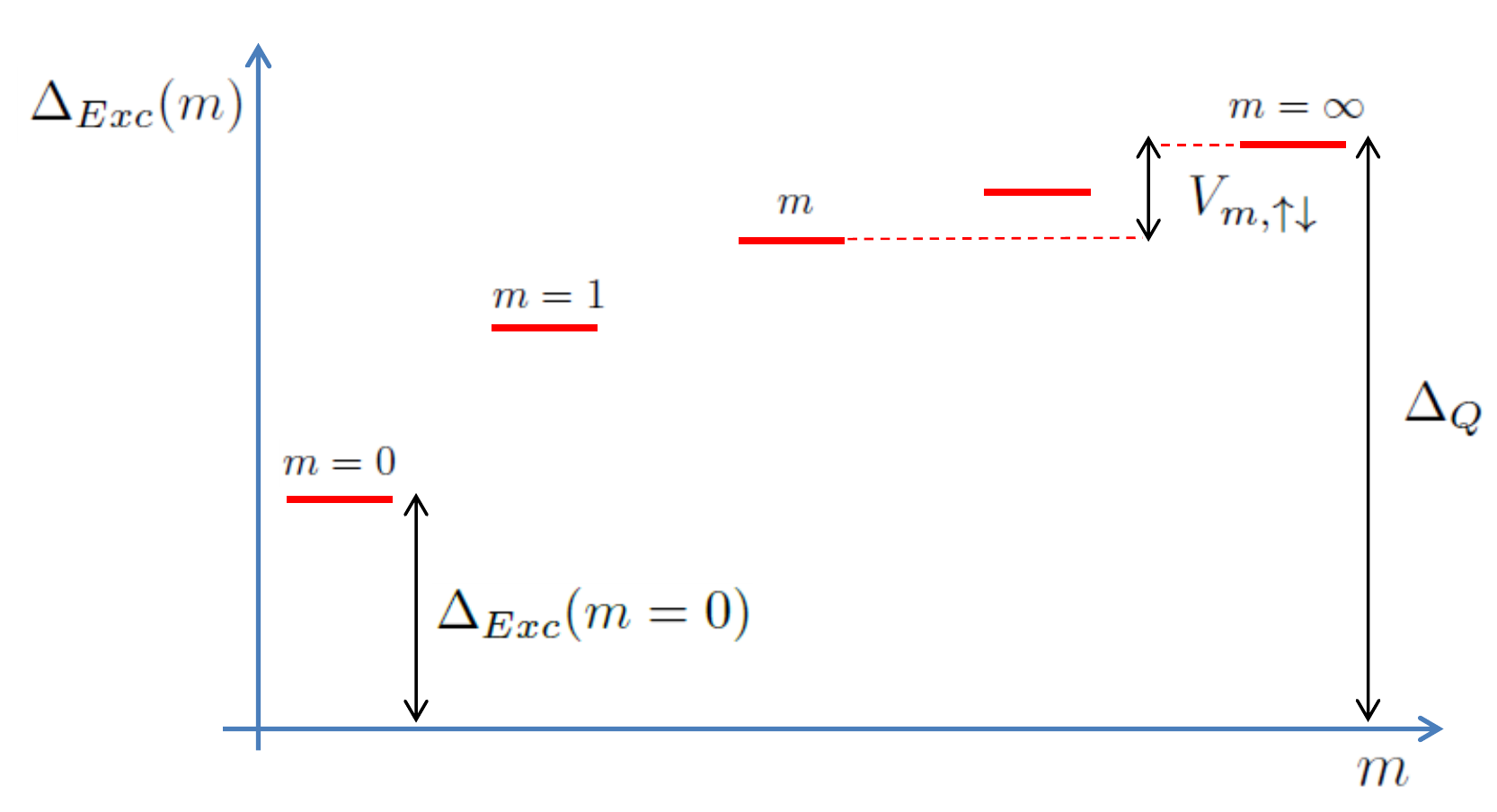}
\caption{Spectrum of exciton states on the Ising Chern magnet. $m$ labels the relative angular momentum. If the exciton gap $\Delta_{Exc}$ vanishes the state becomes unstable.}
\label{ESp}
\end{center}
\end{figure}

Upon performing this transformation on Eq.~\eqref{V}, the Hamiltonian becomes:
\begin{equation} \label{HP}
\begin{split}
PVP^{\dagger}&=E_{0}+\bar{V}+\epsilon_{\uparrow}N_{ \uparrow}+\epsilon_{\downarrow}N_{ \downarrow}, \\
\bar{V}&=\sum_{i<j}V_{\uparrow \uparrow}(\mathbf{r_i}-\mathbf{r_j})\delta_{\sigma_{z_i},\sigma_{z_j}}-V_{\uparrow \downarrow}(\mathbf{r_i}-\mathbf{r_j})\delta_{\sigma_{z_i},-\sigma_{z_j}}, \\
\epsilon_{\uparrow}&=\int \frac{d^2 q}{(2 \pi)^2} |F(\mathbf{q})|^2 V_{\uparrow \uparrow}(\mathbf{q})-\frac{V_{\uparrow \uparrow}(\mathbf{0})}{2 \pi l^2}, \\
\epsilon_{\downarrow}&= \frac{V_{\uparrow \downarrow}(\mathbf{0})}{2 \pi l^2} \\
\end{split}
\end{equation}
Here $E_{0}$ is the energy of the Ising Chern Magnet and $V_{\sigma \sigma'}(\mathbf{q})$ is the Fourier transform of $V_{\sigma \sigma'}(\mathbf{r})$. We have kept global Hartree terms, $V_{\sigma \sigma'}(\mathbf{q}=0)$, but these could be absent if a neutralizing background is assumed. Notice that the role of total particle number, $N = N_{\uparrow} + N_{\downarrow}$, and valley polarization, $Sz = N_{\uparrow} - N_{\downarrow}$, are swapped in the quantum Hall picture relative to the topological insulator picture:
\begin{equation}\label{26}
PNP^{\dagger}=N_{\phi}-S_{z}, \ PS_{z}P^{\dagger}=N_{\phi}-N.
\end{equation}
If a state has valley occupation  in the physical picture $ (N_{\uparrow},N_{\downarrow})$, we will denote its occupation in the quantum Hall picture by $(N_{\uparrow},N_{\downarrow})_{P}=(N_{\phi}-N_{\uparrow},N_{\downarrow})$. The physical quasi-electron and quasi-hole excitations have respectively $ (N_{\uparrow},N_{\downarrow})=(N_{\phi},1),(N_{\phi}-1,0)$, or $(N_{\uparrow},N_{\downarrow})_{P}=(0,1),(1,0)$. Their energies relative to the Ising Chern magnet can be read from Eq.~\eqref{HP} and are respectively $\epsilon_{\uparrow},\epsilon_{\downarrow}$. Therefore, the bulk charge gap of the Ising Chern magnet is:
\begin{equation}
\Delta_{Q}=\epsilon_{\uparrow}+\epsilon_{\downarrow}.
\end{equation}
The stability of the Ising Chern magnet requires $\Delta_{Q}>0$. When $V_{\uparrow \uparrow}(\mathbf{r_i}-\mathbf{r_j})$ is chosen to be a delta-function $\epsilon_{\uparrow}=0$  (since Hartree and exchange exactly cancel each other, reflecting the inability of electrons to be at the same position), and $\epsilon_{\downarrow}>0$ for repulsive interactions, and thus $\Delta_{Q}>0$, in agreement with the more general arguments of the previous section. Notice also that when the spatial average of the inter and intra-valley interactions is the same, namely $V_{\uparrow \uparrow}(\mathbf{q}=0)=V_{\uparrow \downarrow}(\mathbf{q}=0)$ and the inter-valley interaction is repulsive $\int d^2 q |F(\mathbf{q})|^2 V_{\uparrow \uparrow}(\mathbf{q})>0$, then the charge gap is also positive. The latter criterion encompasses the case of pure Coulomb interactions since the neutralizing background demands $V_{\uparrow \uparrow}(\mathbf{q}=0)=V_{\uparrow \downarrow}(\mathbf{q}=0)$. A more stringent criterion on the stability of the Ising Chern magnet is obtained by studying its problem of particle-hole excitations. The simplest particle-hole excitation has valley numbers $(N_{\uparrow},N_{\downarrow})=(N_{\phi}-1,1)$ or $(N_{\uparrow},N_{\downarrow})_{P}=(1,1)$. Therefore, in the quantum Hall picture these excitons behave as a pair of charged particles in a magnetic field, and their states can be simply obtained by exhausting the symmetries of the problem, in analogy to how the two-body problem in conventional Landau levels is solved~\cite{Prange90}. These states can be labeled by two integers, a center of mass angular momentum, $M$, and a relative angular momentum, $m$. Their energy depends only on the relative angular momentum, and this defines the notion of an exciton Haldane pseudo-potential. The energy of the excitonic state $\ket{M,m}$ can therefore be read then directly from Eq.~\eqref{HP} and it is:
\begin{equation} \label{Exc}
\Delta_{Exc}(m)=\Delta_{Q}-V_{m, \uparrow \downarrow},
\end{equation}
where $V_{m, \uparrow \downarrow}$ is the Haldane pseudo-potential associated with the inter-valley interaction:
\begin{equation} \label{Vhal}
V_{m, \uparrow \downarrow}=\int_{0}^{\infty} q dq V_{\uparrow \downarrow}(q) L_{m}(q^2)e^{-q^2}.
\end{equation}
$V_{m, \uparrow \downarrow}$ can be interpreted as the exciton binding energy which measures the attraction of the electron and the hole as depicted in  Fig.~\ref{ESp}. The stability of the Ising Chern magnet also requires the exciton energy to be positive $\Delta_{Exc}(m)$, namely that its binding energy is smaller than the charge gap $V_{m, \uparrow \downarrow} \leq \Delta_{Q},\ \forall m $, to prevent spontaneous exciton proliferation on top of the Ising magnet vacuum. A schematic depiction of these energies is shown in Fig.~\ref{ESp}.

\begin{figure}[th!]
\begin{center}
\hspace{-0.4in} \includegraphics[width=3.2in]{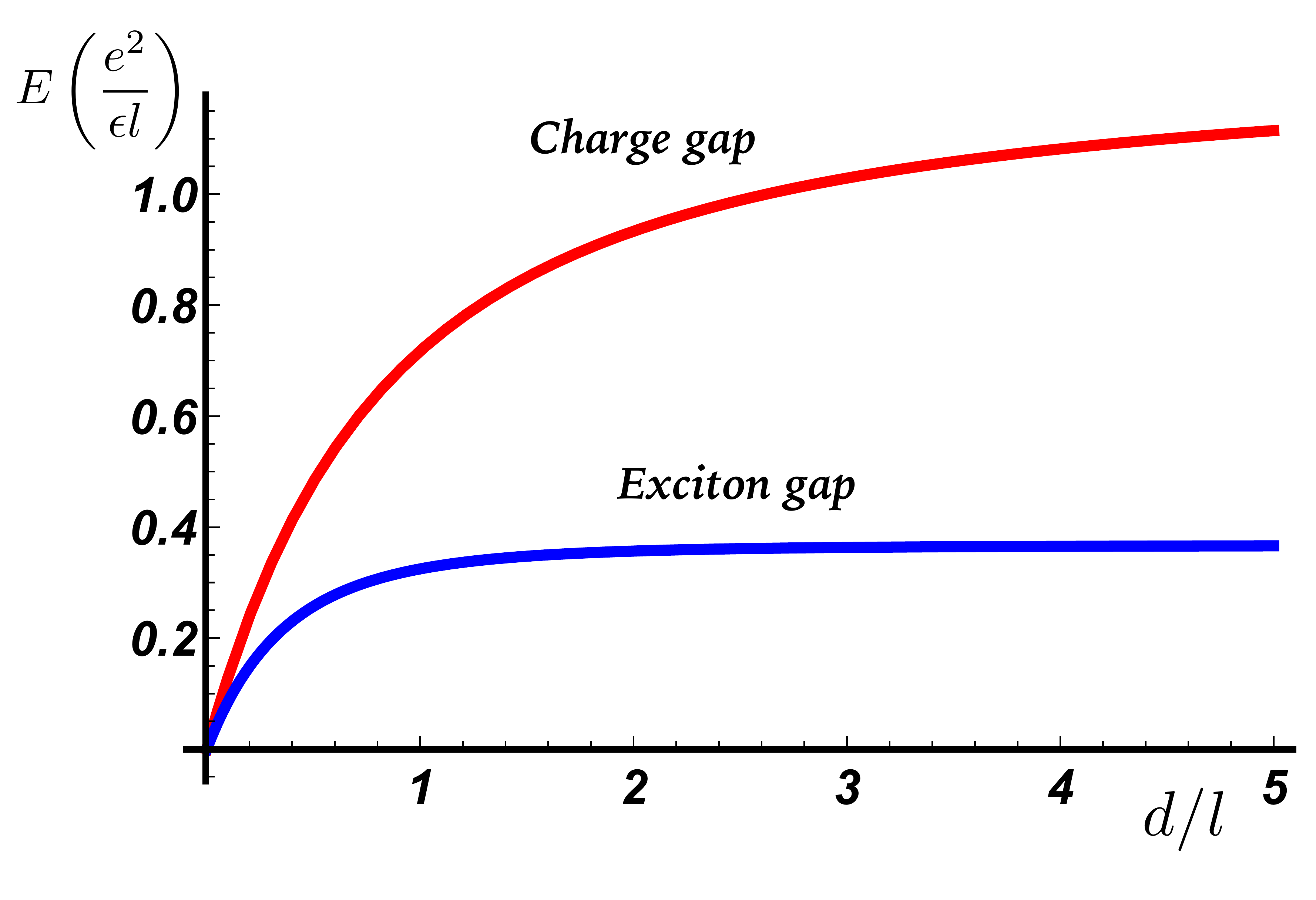}
\caption{Charge and exciton gaps for Coulomb interactions screened by a metallic gate at distance $d/2$.}
\label{Coulombs}
\end{center}
\end{figure}

\begin{figure}[ht!]
\begin{center}
\hspace{-0.325in} \includegraphics[width=3.2in]{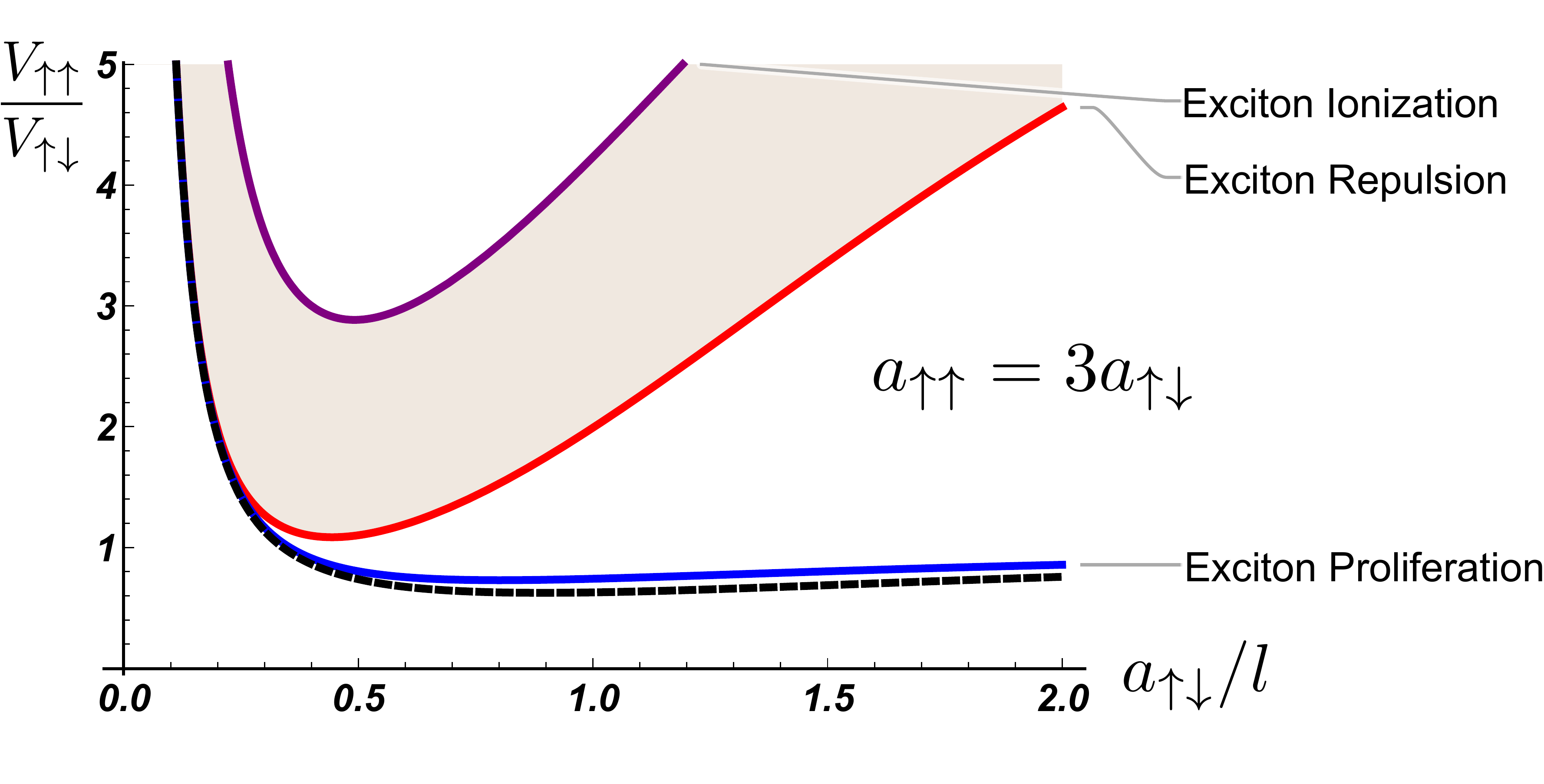}
\caption{Shaded region of optimal conditions for excitonic
Laughlin states for a model of intra-valley and inter-valley repulsions of range $a_{\uparrow \uparrow}=3a_{\uparrow \downarrow}$ and strengths $V_{\uparrow \uparrow}$ and $V_{\uparrow \downarrow}$. At the red line their interaction changes from attractive to repulsive. At the purple line the exciton binding energy is comparable to the inter-exciton interaction. The dashed line indicates the exciton-pair proliferation instability which is lower than the single-exciton proliferation instability, in contrast to the case with $\alpha_{\uparrow \uparrow}=\alpha_{\uparrow \downarrow}$ shown in Fig.\ref{Fig1}(a).}
\label{EH}
\end{center}
\end{figure}

We will discuss now the stability of the Ising Chern magnet against exciton proliferation for two concrete microscopic interactions. For moir\'e super-lattice materials it is likely that the density-density interactions are roughly the same for inter- and intra-valley interactions, since these degrees of freedom are related by a spatially local time reversal symmetry. Bringing a metallic gate near the bilayer produces a modified Coulomb interaction of the form: 
\begin{equation}
V_{\uparrow \uparrow}(r)=V_{\uparrow \downarrow}(r)=\frac{e^2}{\epsilon r}-\frac{e^2}{\epsilon\sqrt{r^2+d^2}},
\end{equation}
where $d$ is twice the distance to the metal gate (distance to the image charges). This strategy has been successfully employed recently to induce non-trivial changes in the physics of moir\'e super-lattice materials~\cite{Stepanov19}. The charge and the exciton gap for these interactions are given by:
\begin{equation}
\begin{split}
\Delta_{Q}&=\sqrt{\frac{\pi}{2}}\frac{e^2}{\epsilon l}\bigg(1-e^{2 \frac{d^2}{l^2}}Erfc \Big[\sqrt{2}\frac{d}{l}\Big]\bigg), \\
\Delta_{Exc}(m=0)&=\frac{e^2 \sqrt{\pi}}{\sqrt{2} \epsilon l}\Bigg[-\frac{1}{\sqrt{2}}\bigg(1-e^{ \frac{d^2}{l^2}}Erfc\Big[\frac{d}{l}\Big]\bigg) \\
&+\bigg(1-e^{2\frac{d^2}{l^2}}Erfc\Big[\sqrt{2}\frac{d}{l}\Big]\bigg)\Bigg]. \\
\end{split}
\end{equation}
These quantities are depicted in Fig.~\ref{Coulombs}
and therefore we see that although these quantities decrease as the metallic gate becomes closer to the system, they remain positive and reveal no instability of the Ising Chern magnet for any finite $d$. This provides further evidence of the robust stability of the Ising Chern magnet, and is in broad agreement with experiments that have advocated its observation under diverse circumstances \cite{Stepanov19,Serlin20}.
However, there is a possibility, as we will see, that certain correlated states that we will call excitonic Laughlin states become viable energetic competitors to the Ising Chern magnet after short distance valley-dependent modifications to the Coulomb interactions are added. With this motivation, we consider a toy model of finite ranged interactions modeled as Gaussians:
\begin{equation} \label{G}
V_{\uparrow \uparrow}(\mathbf{r})=\frac{1}{2\pi a^2_{ \uparrow \uparrow}} V_{\uparrow \uparrow}e^{\frac{-r^2}{2a^2_{\uparrow \uparrow}}}, \ V_{\uparrow \downarrow}(\mathbf{r})=\frac{1}{2\pi a^2_{ \uparrow \downarrow}} V_{\uparrow \downarrow}e^{\frac{-r^2}{2a^2_{\uparrow \downarrow}}}
\end{equation}
We restrict to the case of repulsive interactions: $V_{\uparrow \uparrow},V_{\uparrow \downarrow} \geq 0$. The charge gap and the exciton gap in this case are:
\begin{equation}\label{ExcG}
\begin{split}
\Delta_{Q}&=\frac{V_{\uparrow \downarrow}}{2 \pi}-\frac{V_{\uparrow \uparrow}}{2 \pi}\bigg(\frac{\alpha^2_{\uparrow \uparrow}}{l^2+\alpha^2_{\uparrow \uparrow}}\bigg), \\
\Delta_{Exc}(m=0)&=\frac{V_{\uparrow \downarrow}}{2 \pi}\bigg(\frac{l^2+\alpha^2_{\uparrow \uparrow}}{2l^2+\alpha^2_{\uparrow \uparrow}}\bigg)-\frac{V_{\uparrow \uparrow}}{2 \pi}\bigg(\frac{\alpha^2_{\uparrow \uparrow}}{l^2+\alpha^2_{\uparrow \uparrow}} \bigg). \\
\end{split}
\end{equation}
Fig.~\ref{Fig1}(a) and Fig.~\ref{EH} depict the boundary where these gaps vanish indicating an instability of the Ising Chern magnet, where a different state takes over as the ground state. Although this is a sufficient criterion for the instability of the Ising Chern magnet, it is not necessary. As we will see explicitly in the next section multi-exciton processes can sometimes destabilize  the state before the single exciton instability appears.

\section{Exciton interactions and Excitonic Laughlin states} The fact that excitons in topological insulating bands behave as charged particles in a magnetic field leads to the natural possibility that once the Ising Chern magnet is no longer the ground state and excitons proliferate they could form correlated states that appear typically in partially filled Landau levels. A natural possibility are states with broken translational symmetry which have been considered in \cite{Bultinck19,Chatterjee19} and which we will discuss in a forthcoming publication~\cite{Stefanidis20}. Here we would like to discuss another possibility, namely that of excitonic Laughlin-type states. Once they proliferate, it is reasonable to expect the excitons to form bosonic Laughlin states, provided the following two criteria are satisfied: (a) the excitons are well bound, namely the exciton ionization energy is large compared to the typical inter-exciton interactions and (b) the excitons have repulsive interactions. Under these conditions, a Laughlin state of $N_{X}$ excitons can form at $(N_{\uparrow},N_{\downarrow})=(N_{\phi}-N_{X},N_{X})$ or $(N_{\uparrow},N_{\downarrow})_{P}=(N_{X},N_{X})$ provided that the exciton filling factor, defined as $\nu_{X}=N_{X}/2 N_{\phi}$, satisfies:
\begin{equation} \label{nu}
\nu_{X}= \frac{1}{2 m}, \ m=2,3,...
\end{equation}
This exciton filling factor is $1/4$ of the filling factor in the quantum Hall picture defined from Eq.~\eqref{P} and discussed in section \ref{Excitons and stability of Ising Chern magnets}. This is because the exciton number equals half the number of particles in this picture and the excitons experience twice the magnetic field strength of the particles, and hence twice the number of effective flux quanta. In defining the Laughlin states of excitons and the exciton filling factors one has in mind a picture of tightly bound excitons. However, we would like to mention that the limit of tightly bound excitons and the limit defining the Landau level projection have certain degree of conflict. For example, if the excitons have filling $\nu_{X}=1/2$, we would conclude that particles would have valley numbers $(N_{\uparrow},N_{\downarrow})=(0,N_{\phi})$, namely it would correspond to the Ising Chern magnet with opposite polarization to our reference vacuum, $(N_{\uparrow},N_{\downarrow})=(N_{\phi},0)$. Clearly it is impossible to have a topologically ordered state at such filling if Landau levels are infinitely far away in energy. We believe, however, that the universal properties of tightly bound excitonic states can be recovered under strict Landau level projection for states with exciton fillings $0<\nu_{X}<1/2$ and this is the reason we have excluded $m=1$ as a possibility in Eq.~\eqref{nu}. 

Now, in order to verify the two criteria that make amenable the appearance of excitonic Laughlin states, we are led to consider the problem of short-distance inter-exciton interactions. For concreteness we will focus on the model with Gaussian interactions introduced in Eq.~\eqref{G}. To study the short distance interactions we consider s four-particle state with valley numbers $(N_{\uparrow},N_{\downarrow})=(N_{\phi}-2,2)$ or $(N_{\uparrow},N_{\downarrow})_{P}=(2,2)$, corresponding to two-excitons, $N_{X}=2$,  that describes the case of closest approach between these particles allowed by the Pauli exclusion principle:
\begin{equation}
\Psi_{2X}=(z^{\uparrow}_{1}-z^{\uparrow}_{2})(z^{\downarrow}_{3}- z^{\downarrow}_{4})e^{-\frac{\sum_{i}|z_i|^2}{4 l^2}}\ket{{\uparrow}_1 {\uparrow}_2 {\downarrow}_3 {\downarrow}_4},
\end{equation}
here the state is written in the symmetric gauge in the quantum Hall picture defined in Eq.~\eqref{P} and normalization and full antisymmetrization are implicit. In the physical picture, the $\uparrow$ particles are the holes in the Ising magnet vacuum and the $\downarrow$ particles are added to the empty valley. This state can be viewed as a quantum Hall droplet of $\nu=2$ constructed on top of the Ising Chern magnet. Because this is the most compact two-exciton state, the energy of this state can be viewed as characterizing the analogue of the Haldane pseudo-potential $V_{2}$ for the excitons, and due to their underlying fermionic constituents the excitons behave as hard-core bosons with infinite $V_{0}$. States in which the excitons are farther apart are expected to have lower interaction energy, because their interaction decays with distance for simple models of microscopic interactions such as the Gaussian one. More specifically, the energy of $\Psi_{2X}$ measured relative to the Ising Chern magnet, can be decomposed into exciton gap and exciton interaction parts as follows:
\begin{equation} \label{E2X}
E_{2X}=\expval{V_{P}}{\Psi}-E_{0}=2 \Delta_{X}+V_{2X},
\end{equation}
where $V_{P}$ is given in Eq.~\eqref{HP}, $E_{0}$ is the energy of the Ising Chern Magnet, and $\Delta_{X}$ is the single exciton gap defined in Eq.~\eqref{Exc} and given in Eq.~\eqref{ExcG} for the Gaussian model. For this Gaussian model the explicit expression for the exciton interaction is found to be:
\begin{equation} \label{ExInt}
\begin{split}
V_{2X}&=\frac{V_{\uparrow \downarrow}}{\pi}\bigg[\frac{1}{2}\frac{1}{2+\alpha^2_{\uparrow \downarrow}}-\frac{1+\alpha^2_{\uparrow \downarrow}}{2+\alpha^2_{\uparrow \downarrow}}-\frac{1}{2}\frac{2+2\alpha^2_{\uparrow \downarrow}+\alpha^4_{\uparrow \downarrow}}{(2+\alpha^2_{\uparrow \downarrow})^3}\bigg] \\
&+\frac{V_{\uparrow \uparrow}}{\pi}\frac{\alpha^2_{\uparrow \uparrow}}{(2+\alpha^2_{\uparrow \uparrow})^2}. \\
\end{split}
\end{equation}
The boundary separating the region of effective inter-exciton attraction and repulsion is shown as a blue line in Fig.~\ref{Fig1}(a) for the case in which inter-valley and intra-valley interactions have the same range ($a_{\uparrow \uparrow}=a_{\uparrow \downarrow}$) and as a red line in Fig.~\ref{EH} for the case in which the ranges are different ($a_{\uparrow \uparrow}=3a_{\uparrow \downarrow}$). We have also added a purple line that qualitatively determines the region where the excitons are strongly bound from those in which they are not by determining when the exciton binding energy given by $V_{m=0, \uparrow \downarrow}$ in Eq.~\eqref{Vhal} becomes equal to the magnitude of the exciton interaction $V_{2X}$ from Eq.~\eqref{ExInt}. The intersection of the regions in which the Ising magnet is unstable, the excitons are strongly bound, and the excitons have repulsive interactions, is expected to be a fertile ground for the appearance of excitonic Laughlin states and it is shown as a shaded region in Fig.~\ref{Fig1}(a) and Fig.~\ref{EH}.

Importantly, the stability of the Ising Chern magnet also demands $E_{2X}>0$, otherwise the state would be unstable to exciton-pair proliferation processes. Fig.~\ref{EH} also shows the line of exciton pair proliferation $E_{2X}=0$, as a dashed black line. We have found that when intra and intervalley interactions have different range $\alpha_{\uparrow \uparrow} \neq \alpha_{\uparrow \downarrow}$ the boundary of the single exciton proliferation instability, $\Delta_{2X}=0$, generally differs from the boundary of the exciton pair proliferation instability, $E_{2X}=0$, and, in fact, the exciton pair production instability provides a more stringent criterion for the stability of the Ising Chern magnet as seen in Fig.~\ref{EH}. However, when interactions have the same range, the single exciton and exciton-pair proliferation lines coincide, and this is the reason there is no dashed line in Fig.~\ref{Fig1}(a).

Let us now discuss the properties of the excitonic Laughlin states. These states break spontaneously the time reversal symmetry of the topological insulator band. For the cases with $m \geq 3$ in Eq.~\eqref{nu}, this is evident because they have a net valley polarization:
\begin{equation}
(N_{\uparrow},N_{\downarrow})=N_{\phi}\bigg(1-\frac{1}{m},\frac{1}{m}\bigg).
\end{equation}
These states with $m \geq 3$ will have therefore some amount of orbital magnetism although it will tend to be smaller than that of the Ising Chern magnet. Notably, the state corresponding to $m = 2$ in Eq.~\eqref{E2X} which corresponds to exciton filling $\nu_{X}=1/4$ has equal occupation of both valleys $(N_{\uparrow},N_{\downarrow})=N_{\phi}(1/2,1/2)$. However, in spite of having zero valley polarization, this state breaks time reversal symmetry spontaneously as we will demonstrate next.

To show that this state breaks time reversal symmetry we begin by noting that the physical time reversal operator, $T$ from Eq.~\eqref{T}, acts as an anti-unitary particle-hole conjugation in the quantum Hall picture defined by the transformation $P$ from Eq.~\eqref{P}. Namely, the operator $T_{P}=PTP^{-1}$ acts as:
\begin{equation}
T_{P}c^{\dagger}_{m \uparrow}T^{-1}_{P}=c_{m \downarrow}, \ T_{P}c^{\dagger}_{m \downarrow}T^{-1}_{P}=c_{m \uparrow}.
\end{equation}
Once combined with the charge conjugation defined in Eq.~\eqref{C}, this symmetry is equivalent to the anti-unitary particle-hole symmetry that plays an important role in conventional quantum Hall bilayers~\cite{Sodemann17}. Now we consider the $\nu_{X}=1/4$ excitonic Laughlin state placed in the surface of the sphere~\cite{Haldane83}. The sphere induces a finite size shift in the proportion between particles and fluxes, and this shift is a topological invariant of the state~\cite{Wen92}. For a tightly bound $\nu_{X}=1/4$  excitonic Laughlin state this proportion is:
\begin{equation}
2 N_{\phi}=4 (N_{X}-1).
\end{equation}
Where we are using the fact that tightly bound excitons experience twice the magnetic flux of the particles. Therefore the exciton number would be $N_{X}=(N_{\phi}+2)/2$, and thus requires an even number of flux quanta to be realized in the sphere. On the other hand, because the number of single particle orbitals in the sphere is $N_{\phi}+1$ a two-valley fermion state that is particle-hole invariant (i.e. time reversal invariant in the physical picture) must satisfy $(N_{\uparrow},N_{\downarrow})_{P}=((N_{\phi}+1)/2,(N_{\phi}+1)/2)$ and thus would have an exciton number, $N_{X}=(N_{\phi}+1)/2$, and thus requires an odd number of flux quanta to be realized in the sphere. The inconsistency between these conditions implies that the $\nu_{X}=1/4$ excitonic Laughlin state breaks spontaneously time reversal symmetry, and therefore the state constructed by adding excitons to valley $\uparrow$ is distinct from the one obtained by adding excitons to valley $\downarrow$, even though both have zero valley polarization.

We will now describe the nature of bulk excitations of the Excitonic Laughlin states. These will have fractionalized quasi-particles with exciton number quantized in units of $\delta N_{X}=1/2m$. Therefore the fractionalized excitations will be charge neutral but contain an excess of valley numbers of the form $\delta N_{\uparrow}=- \delta N_{\downarrow}=q/2m, \ q \in \mathbb{Z}$. As it is customary with Laughlin states, quasihole and quasi-electrons will have a different spatial profile. The quasi-holes will typically have smaller spatial extend and their core will essentially be a small bubble of fully polarized the parent Ising Chern magnet state. The quasi-electrons will typically be more spatially spread, and at its core will try to recover the opposite valley polarization. An illustration appears in Fig.~\ref{Fig1}. Due to their local valley polarization, these quasi-particles will have therefore a non-trivial profile of local orbital magnetization relative to the background, specially in the $m=2$ Laughlin state, since the background has vanishing valley polarization and this might allow the detection of the states should they occur in moiré superlattice materials. On the other hand, charged quasiparticles will be ordinary quasi-electrons and quasi-holes with integer quantized charges.

Finally, although the edge states can be fairly complex in realistic systems, at least in the limit of strong exciton binding, one expects the excitonic Laughlin states to have the same charge edge channel as the parent Ising magnet on which it is constructed, and therefore the same charge Hall conductivity $\sigma_{xy}=\pm e^2/h$. This is because the excitons are added to a filling factor one parent state, but they are charge neutral particles and one expects them to have a charged edge mode together with a counter-propagating neutral mode, as depicted in Fig.~\ref{Fig1}(b). This property makes it hard to distinguish the excitonic Laughlin states from ordinary Ising Chern Insulators in usual charge transport experiments. In this sense one could say that it is hard to rule out that they might already be present even in the experiments so far reporting the occurrence of spontaneous anomalous Hall effect \cite{Serlin20}.

\section{Summary and Discussion} We have studied the few- and many-body physics of ideal maximally symmetric topological insulator flat bands with repulsive interactions. We have found that charged particle-particle pairs behave in a similar fashion to neutral particle-hole pairs in conventional Landau levels, displaying a form of locking of relative distance and center of mass momentum degrees of freedom. Conversely, neutral particle-hole pairs in topological insulator flat bands behave similarly to charged particle-particle pairs in Landau levels, having a flat dispersion for their center of mass degrees of freedom that exhibit an analogue of cyclotron motion while their relative angular momentum allows to define a notion of exciton Haldane pseudo-potentials. 

We have constructed ideal Hamiltonians for which it can be rigorously argued that the ground state at total filling $1$ is the spontaneously polarized Ising Chern magnet and studied its stability to single exciton and exciton-pair proliferation processes. We have also studied the short range inter-exciton interactions, and demonstrated that once excitons proliferate they are repulsive for a model of short range interactions in which intra- and inter-valley interactions have the same range as shown in Fig.~\ref{Fig1}(a). Taking this range to be comparable to the inter-particle distance ($a\sim l$), we have found that the Ising Chern magnet is no longer the ground state when the intra-valley repulsions are about $30\%$ larger than the intra-valley repulsions. We have argued that Laughlin states of excitons are energetically competetive ground states once the Ising Chern magnet is destabilized. Remarkably, these states display only valley fractionalized charge-neutral quasiparticles, namely, the charge of all quasiparticles is quantized in units of the electron charge. These excitonic Laughlin states have a charge Hall conductivity that is identical to Ising Chern magnet, $\sigma_{xy}= \pm e^2/h$, and thus are hard to distinguish from them in conventional charge transport experiments. In particular, the most compact excitonic Laughlin state is an analogue of the $\nu=1/4$ bosonic Laughlin state, and has no valley polarization in spite of breaking spontaneously the time reversal symmetry with a Hall conductivity $\sigma_{xy}= \pm e^2/h$. Due to the orbital magnetism of the valley polarized states~\cite{JihangZhu20}, the valley fractionalized quasiparticles of these states could display substantial local orbital magnetic moments, and local magnetometry probes could be used to image these quasi-particles. 

\emph{Note added}---~During the completion of our work, other studies with overlapping ideas and results appeared in \cite{Kwan201,Kwan202}.

\appendix

\section{Absence of exact topological degeneracy for unpolarized states} \label{Appendix}
We will now demonstrate that in contrast to the ordinary quantum Hall Landau levels, there is no generic exact topological degeneracy for time reversal invariant states in the Torus, by extending the classic analysis of Haldane~\cite{Haldane85} to our topological insulating flat bands. To place the system in a torus requires the ability to simultaneously diagonalize translations for two non-collinear vectors $\mathbf{L_x}$,$\mathbf{L_y}$  that define the principal axes of the torus, and thus the system must enclose an integer number flux quanta: $BA=2\pi N_{\phi}$. Similar to what happens in ordinary Landau levels, the torus induces a weak breaking of translational symmetry, in the sense that only a finite subgroup of the magnetic translation algebra is compatible with any given torus with a specific choice of twist of boundary conditions. More specifically, one can show that only the following subset of single particle translations commutes with the translations $t_{\mathbf{L_x}}$,$t_{{\mathbf{L_y}}}$ defining the torus:
\begin{equation}
t(n,m)=t \Big(\frac{n}{N_{\phi}}\mathbf{L_{x}}+\frac{m}{N_{\phi}}\mathbf{L_{y}}\Big),n,m \in \mathbb{Z}.
\end{equation}
The many-body or center-of-mass translation operators are defined as:
\begin{equation}\label{11}
T(n,m)=\prod_i^{N_{e}} t_i(n,m),
\end{equation}
where $i$ labels the particles in the system. The smallest allowed translations of the center of mass are $T(1,0)$ and $T(0,1)$. Both of these operators commute with the Hamiltonian, but do not necessarily commute with each other, but instead obey the following algebra:
\begin{equation}
T(1,0)T(0,1)=e^{i2\pi\frac{\sum_i \sigma^z_{i}}{N_{\phi}}}T(0,1)T(1,0).
\end{equation}
Thus we are lead to introduce the notion of the polarization filling factor:
\begin{equation}
\nu_{z}=\frac{\mid \sum_{i=1}^{N_e} \sigma^{z}_i \mid}{N_{\phi}}.
\end{equation}
This is a rational number that in general differs from the ordinary filling factor $\nu=N_e/N_{\phi}$, and which can be decomposed as $\nu_{z}=p_z/q_z$, with $p_z$,$q_z$ relative primes. Then, one can show that algebra of Eq.~\eqref{11} implies an exact $q_z$-fold degeneracy of all the eigenstates of the Hamiltonian, and in particular its ground state. All other translation operators with larger $n,m$ can be constructed as powers of the two smallest translations $T(1,0),T(0,1)$ and therefore are not independent and do not lead to extra degeneracies. This generalizes the criterion by Haldane that guarantees a $q$-fold degeneracy for a state in a Landau level with filling factor $\nu=p/q$~\cite{Haldane85}.
\par States that are valley un-polarized have $\nu_z=p_z=0$, and therefore do not have exact degeneracies enforced by the many-body translation algebra. In particular, time-reversal invariant states are a subset of valley un-polarized states (any time-reversal invariant is valley unpolarized, but the converse is not necessarily true). Valley unpolarized states have necesarilly an even number of particles and therefore the many body time reversal symmetry squares to $T^2=1$ and does not imply extra degeneracies. Thus, as a corollary, we conclude also that any time reversal invariant state in a partially filled flat TI band has no generic exact topological degeneracies.
This should also hold in a realistic partially filled TI bands, because these can be seen as a problem in which some of the symmetries we have enforced exactly are explicitly broken, and therefore this can only result in splitting of the degeneracies that are present in our situation of ideal flat bands.
\par It is important to emphasize that the lack of exact ground state degeneracy does not imply the absence of topological order, but only that if the latter exists its degeneracy will only appear asymptotically in the thermodynamic limit. We also note that an alternative quick way to recover the results of this supplementary can be achieved by performing a partial particle-hole transformation in one of the flavors as described in the Section \ref{Excitons and stability of Ising Chern magnets} since the valley polarization is mapped into the total filling factor, as shown in Eq.~\eqref{26}.

\bibliographystyle{apsrev4-1}

\begin{thebibliography}{100}
\bibitem{footnote} This implies that, strictly speaking, we don't need to require $V_{\uparrow \uparrow}>0$ for the current argument.
\end{thebibliography}

\end{document}